# Neurobiology and Changing Ecosystems: toward understanding the impact of anthropogenic influences on neurons and circuits

Angie Michaiel and Amy Bernard
The Kavli Foundation, United States

## Abstract

Rapid anthropogenic environmental changes, including those due to habitat contamination, degradation, and climate change, have far-reaching effects on biological systems that may outpace animals' adaptive responses (Radchuk et al., 2019). Neurobiological systems mediate interactions between animals and their environments and evolved over millions of years to detect and respond to change. To gain an understanding of the adaptive capacity of nervous systems given an unprecedented pace of environmental change, mechanisms of physiology and behavior at the cellular and biophysical level must be examined. While behavioral changes resulting from anthropogenic activity are becoming increasingly described, identification and examination of the cellular, molecular, and circuit-level processes underlying those changes are profoundly underexplored. Hence, the field of neuroscience lacks predictive frameworks to describe which neurobiological systems may be resilient or vulnerable to rapidly changing ecosystems, or what modes of adaptation are represented in our natural world. In this review, we highlight examples of animal behavior modification and corresponding nervous system adaptation in response to rapid environmental change. The underlying cellular, molecular, and circuit-level component processes underlying these behaviors are not known and emphasize the unmet need for rigorous scientific enquiry into the neurobiology of changing ecosystems.

## Introduction

Nervous systems evolved across hundreds of millions of years to allow organisms to sense and respond to external environmental changes. Being receptive and responsive, yet flexible, are features that lie at the core of the remarkable adaptive capabilities of the brain. Environments and organisms are inseparable. Evolution shapes the nervous system through experience with a dynamic range of environmental possibilities. However, rapid, human-induced environmental change may challenge the range of experience in evolutionary history as a function of the adaptive clock mechanisms imposed by genetics. How resilient are neural systems in the face of unanticipated and potentially prolonged change? What are the rules that govern whether an animal will adapt within its ecological niche, or attempt to find another one? Nervous systems mediate the interactions between organisms and their environments, and thus, understanding their response to anthropogenic environmental change is vital for the ability to predict *what* changes may transpire with rapid environmental perturbation, and *how* neuronal systems may adapt to – or fail to adapt – to those changes.

Anthropogenic pressures, including but not limited to, dramatic increases in greenhouse gases and subsequent climate instability, habitat depletion, and pollution have emerged only recently in our planet's history. The most optimistic future climate change scenario, assuming nationally determined contributions are implemented by 2030, predicts warming of an average of 2.4ºC by 2100 (Kemp et al., 2022). Following our current trajectory, warming will increase by an average of 3ºC. In more devastating scenarios, positive feedback loops in the carbon cycle may prevent stabilization of climate and trigger sudden and irreversible global warming in unpredictable and catastrophic ways (Steffen et al., 2018). Continuing our current trajectory is predicted to lead to an estimated extinction of one-third of all animal and plant species globally by 2070 (Román-Palacios and Wiens, 2020).

Behavioral responses of individual animals and populations to anthropogenic environmental influence are becoming increasingly well described (Tuomainen and Candolin, 2011; Candolin and Wong, 2012; Goumas et al., 2020; Gunn et al., 2022). Connecting phenomenological observations of anthropogenic influence on animal behavior to causative cellular, molecular, and circuit-level processes, however, is an open area that is ripe for inquiry. For example, certain pollutants reduce the efficiency of escape responses in fish (Weis et al., 1999, 2001; Lürling and Scheffer, 2007). Likewise, the genes, cells, and circuits that mediate escape responses are known, but the connection to the phenomenological observation does not yet exist. How do environmental pollutants modify known behavioral response mechanisms at all these levels of analysis? The advent of tools to target, monitor, and manipulate genes, cells, and circuits has yielded an immense understanding of basic neurobiology that can now be integrated with ethologically inspired approaches and questions.

Responses to anthropogenic impacts may originate or affect neural systems at multiple levels, from changes in gene expression and neural plasticity all the way to cognition, each of which likely operate under different timescales and with varying robustness to change. Take for example cases of noise pollution in which machine sounds interfere in the same frequency range that certain species use for communication. If neural circuit activity and behavior are adaptive, frequencies of sounds used to communicate may shift (though this may create new pressures). If neural circuit plasticity does not appropriately occur, perhaps there may instead be selection for different frequency communication schemes spanning across multiple generations. Environmental changes may be too rapid for some species to survive. Species responses will also be variable, ultimately altering ecosystems in potentially irreversible ways.

Unprecedented anthropogenic change also brings forth scientific questions that would not have been imagined decades ago and would not have been approachable without the tools and domain knowledge of the present day. Anthropogenic influence on ecosystems is relatively new; we were unable to observe these phenomena because they simply didn't exist or were unmeasurable with the available methods and technologies. Similarly, we didn't have the domain knowledge within neuroscience (and we still do not) that would allow us access into the question space. How could we predict how noise pollution impacts auditory system development without an understanding of normal development? As studying disease can provide useful domain knowledge about basic biological processes, studying neural systems under duress may help to explicate both normal and disrupted neural function. Now is the time to combine our understanding of anthropogenic influence on ecosystems with our understanding of neuroscience domain knowledge or utilize current environmental pressures to build additional domain knowledge; now we have an entry point.

Pursuing the mechanistic foundations of some phenomenological observations with the same spatial and temporal specificity that can be achieved in well-controlled lab settings may be unfeasible, these observations have broad implications on animal behavior, sensory plasticity, contemporary evolution, and neurophysiological function. Researchers are, however, addressing questions that can feasibly be approached in the lab or at the very least, inspire new tools and conceptual frameworks. This review describes selected examples of anthropogenic influence on climate with corresponding impacts to animal behavior and explores underlying neural mechanisms. More specifically, we explore stressors imposed by ocean acidification, rises in temperature, and changes in sensory environments and how they may impact neural function (Figure 1). We view these studies as the vanguard in a new and transformative field of neuroscientific research that has the potential to transform our comprehension of the natural world and of nervous system function itself.

## Ocean Acidification

Rising atmospheric $CO_2$ from greenhouse gas emissions generates an uptake in carbonic acid within surface ocean levels, resulting in lower pH marine environments. This process, known as ocean acidification (OA) has not been observed at the current rate in the past 300 million years of inferred geologic history and is predicted to result in a global reduction of an estimated 0.7 pH units by the year 2300 (Caldeira and Wickett, 2003). OA has become a major focus of climate change study due to its link to physiological impairments on developmental, metabolic, immune, and respiratory processes (Esbaugh, 2018).

Influences of variable pH on neural excitability have long been studied, though not entirely through the lens of anthropogenic OA. Due to the pH-sensitive nature of membrane proteins expressed in neurons, namely ion channels, neurotransmitter receptors, transporters, and pumps, neural excitability is modified in acidified conditions (Figure 2; for a complete review see (Ruffin et al., 2014). For example, acidified conditions change conductance and gating properties of voltage-gated sodium, calcium, and potassium channels (Tombaugh and Somjen, 1996) and also reduce current through AMPA and NMDA receptors (McDonald et al., 1998); (Giffard et al., 1990). Activity of $Ca^{2+}$-ATPase pumps increases (Pick and Karlish, 1982; Irwin et al., 1994; OuYang et al., 1994; Wolosker et al., 1997), as does expression of the $Na/HCO_3$ cotransporter, NBCn1 (Park et al., 2010). These pH-sensitive membrane proteins affect neural excitability through a number of processes: regulating the resting membrane potential, setting thresholds for action potential firing, affecting responsiveness to receptor agonists and antagonists, setting the length of refractory periods, and synchronizing neuronal network activity (Ruffin et al., 2014).

The complexity of potential downstream effects of any combination of acidosis on these membrane proteins is difficult to predict, however, one can reasonably speculate that OA may modify neural excitability in aquatic animals and consequently may influence animal behavior. In fact, numerous experimental studies are beginning to link OA to changes specific behaviors such as foraging (Esbaugh, 2018; Jiahuan et al., 2018; Van Colen et al., 2020); reproductive behavior (Milazzo et al., 2016; Nagelkerken et al., 2021); defense behaviors (reviewed in

Clements and Comeau, 2019; predator-prey interactions (reviewed in (Draper and Weissburg, 2019); and habitat selection (Wang and Wang, 2021). At this time, there is little clarity on which behavioral changes are the result of behavioral adaptations, perturbations on pH-dependent physiological mechanisms, or direct perturbations on cell and circuit neurobiological processes that underlie behavior.

An alternate model suggests that impacts of OA on behavior of aquatic animals are in part mediated by changes in physiological pH which is thought to indirectly decrease inhibitory neurotransmission. In a set of studies, larval clownfish reared in lower pH conditions (higher $CO_2$) than controls demonstrated altered olfactory discrimination related to predator cues and cues that would allow them to locate suitable habitat settlement sites (Munday et al., 2009; Dixson et al., 2010). More specifically, fish exposed to high CO2 were attracted to predator odors, rather than avoidant. Interestingly, the onset timing of these behavioral alterations was dependent on the magnitude of pH reduction but was reversible following two-day exposure to control seawater (Munday et al., 2010). These results hinted at an unknown $CO_2$-dependent, yet reversible mechanism that altered behavior. Typically, when exposed to high $CO_2$/low pH conditions, known as hypercapnic conditions, fish regulate acid-base relevant ion concentrations, namely bicarbonate [$HCO_3^-$] and chloride [$Cl^-$], in order to maintain blood and tissue pH (Ishimatsu et al., 2008). This results in an upregulation of active $H^+$ excretion and an accumulation of $HCO_3^-$ in physiological fluids, which is proposed to reduce $Cl^-$ in extracellular compartments (Tresguerres and Hamilton, 2017). The authors hypothesized that changes in $HCO_3^-$ and $Cl^-$ gradients would reverse net ionic influxes through $GABA_A$ channels, switching their function from inhibitory to excitatory.

To test this hypothesis, the experimenters pretreated high $CO_2$ exposed fish with the $GABA_A$ receptor antagonist, gabazine, and observed a restoration in prey avoidance behavior (Nilsson et al., 2012). These $GABA_A$ receptor antagonist studies have now been replicated in multiple marine and freshwater species (with varying results) and posit that OA conditions result in neuronal hyperexcitability that explains observed changes in behavior (reviewed in Tresguerres and Hamilton, 2017).

The $GABA_A$ model stands as one of the few proposed and experimentally tested basic neurobiological mechanisms downstream of anthropogenic environmental changes. However, it is critical to note that ion concentrations have not been directly measured under hypercapnic conditions that arise from OA and that these results are controversial because animals must be in states of extreme, unnaturally acidic conditions for $GABA_A$ channels to reverse the direction of ion flow. This further highlights the need for targeted approaches to understand how OA may impact membrane protein function in neurons as well as how homeostatic mechanisms that regulate physiological pH may impact neural processes. Species-specific neuronal responses to OA are also quite variable (Tresguerres and Hamilton, 2017), suggesting a potential comparative area of study into system variability, adaptation, and resilience.

# Sensory Ecology and Change

Sensory ecology focuses on *how* and *why* animals obtain information from their environments. Natural animal behaviors largely rely on the perception of a diversity of biologically important environmental sensory cues, for example temperature fluctuation and length of daylight prompting the migration of monarch butterflies (Guerra, 2020). Anthropogenic environmental change presents an opportunity to re-examine the central questions of sensory ecology; *if the sensory environment has changed through anthropogenic activity, what is the impact on behavioral systems?* A considerable body of research is focused on sensory ecology related to changing ecosystems, specifically in how noise, light, and chemical and material pollutants impair sensory cue production, transmission, detection, and discrimination, and yet the neurobiological impacts or consequences of change are not well understood (for comprehensive reviews, see (Kelley et al., 2018; Draper and Weissburg, 2019).

Anthropogenic activity may modify or mask the native sensory information that animals encounter. Sound, light, and chemical pollution, for example, reduce the quality of auditory, visual, and chemosensory cues required to guide behavior and results in changes to behavioral strategies, as illustrated in the following examples. Specific neurophysiological mechanisms governing these behavioral adaptations, however, are not yet explored.

## *Auditory Systems*

Animals may alter their behavior in response to increasing anthropogenic noise from sources such as shipping and boating, automobile, sonar, seismic surveys, deep-sea mining, ocean dredging, wind turbines, and numerous other sources (Tuomainen and Candolin, 2011; Kunc et al., 2016; Harding et al., 2019). Typically, humpback whales rely on high-information content vocal communication with conspecifics, but under high ambient noise conditions shift their communication schemes to instead display more surface-generated sounds at the expense of less information-content (Dunlop et al., 2010). When exposed to low-frequency sonar, however, they increase the repetition of phrases in their songs, suggesting that even within the same species, behavioral modulation to different types of noise itself is stimulus-specific and reflects a need to optimize information content fidelity (Miller et al., 2000). Similar findings have been reported in terrestrial species where urbanization generates loud and low-frequency noise (Warren et al., 2006). For example, some bird species compensate for anthropogenic noise by modulating specific components of their song, in a species-specific manner (Lohr et al., 2003; Brumm, 2004; Halfwerk and Slabbekoorn, 2009; Slabbekoorn and Halfwerk, 2009; Zollinger et al., 2011; Ríos-Chelén et al., 2013; Gil et al., 2015).

In addition to potential changes in neurotransmission, ocean acidification also increases ambient noise in the ocean due to the pH dependence of sound absorption in water, specifically within the range of 0.01 – 10 kHz (Fisher and Simmons, 1977; Hester et al., 2008; Ilyina et al., 2009). This frequency-dependent modification is within the hearing sensitivity range of many invertebrates (Albert and Kozlov, 2016), fishes (Ladich and Fay, 2013) and marine mammals (Weilgart, 2007) and may interfere neuronal circuits that mediate echolocation, navigation, mate choice, habitat selection, and other behaviors that depend on acoustic cues in the marine soundscape.

While impacts of anthropogenic noise on animal behavior are well described in both marine and terrestrial ecosystems, much less, if anything, is known about its influence on neurophysiology and sensory processing within the brain. Several studies monitor changes in auditory sensitivity thresholds using auditory evoked potentials (AEPs) and acoustic brainstem responses (ABRs) and find that anthropogenic noise decreases auditory sensitivity across a range of species including fish (Vasconcelos et al., 2007; Codarin et al., 2009), marine mammals (Nachtigall et al., 2004; Lucke et al., 2009), leatherback turtles (Piniak et al., 2012, 2016), and polar bears (Nachtigall et al., 2007). Damage to auditory systems may be through a hearing organ – such as destruction of hair cells due to high frequency exposure – or through the perturbation of the secondary signal in. auditory circuit processing. It remains enigmatic as to how these changes in hearing system function connect to observed behavioral adaptations and how the brain adapts to dynamic changes in signal-to-noise of acoustic signals. What might unravelling these questions teach us about the limits or flexibility of both behavior and sensory processing?

*Visual Systems*

For many animals, vision is a vital sense that drives behaviors, ranging from mate selection to migration to prey capture, and many others (Cronin et al., 2014). Studies of visual ecology related to anthropogenic activity primarily target three sources that effectively mask visual cues: light pollution, changes in water turbidity, and shifts in the ocean depth that animals inhabit (known as bathymetric shifts).

Artificial light has impacts on wildlife as well as human health by altering natural light regimes spatially, temporally, and spectrally (Gaston et al., 2013; Falcón et al., 2020). Light itself is an information source used to regulate metabolic and physiological processes, circadian rhythms and photoperiodism, migratory and orienting behavior, and visual perception, (such as the impacts of artificial lighting and LED by (Falcón et al., 2020). Take for example, sea turtle hatchlings that use lunar illumination and its reflection on water to navigate to sea. There are several accounts of disoriented hatchlings who travel to erroneous locations owning to light pollution, and even those who do successfully reach the sea sometimes return to the shore when shore-based artificial lighting was present (Truscott et al., 2017).

An emerging body of work is examining plasticity of retinal opsin gene expression in reef fish under differing light regimes (Carleton et al., 2020; Musilova et al., 2021; Fogg et al., 2022). Using transcriptomics and qPCR, experimenters have discovered that artificial light increases expression of cone (but not rod) opsin genes sensitive to a broad range of wavelengths. Shifts in opsin expression under differing light conditions were also reversible and highly plastic. Future work will examine if plastic changes in opsin gene expression are maintained with long-term exposure to artificial light.

Run-off from agricultural and septic sources, dredging, and erosion create murky, high-turbidity environments that mask visual cues in water. In one study, turbidity reduced the reactive distance of predators to their prey, making them less efficient predators (Gregory, 1993). In fact, water turbidity is proposed to shape predator-prey interactions of aquatic species by altering visual sensitivity thresholds (Abrahams and Kattenfeld, 1997; Van de Meutter et al., 2005). Using the optomotor response, researchers have measured reduced visual sensitivity as a function of turbidity (Nieman et al., 2018; Suriyampola et al., 2018). How might the fidelity of visual signals and sensitivity to those signals influence behavioral modifications? In simulated turbid environments, western rainbowfish compensated for turbidity by enhancing the vibrance of their color displays to maintain visual communication with conspecifics (Kelley et al., 2012). It will be fascinating to discover how visual signals are utilized by the brain and integrated with internal and ecological demands to generate behavioral modifications, and at what evolutionary rate

Warming and ocean acidification have caused marine animals to inhabit deeper waters (known as bathymetric shifts; (Perry et al., 2005; Pinsky et al., 2013) where optical conditions are poorer in both illumination and spectral breadth. In these scenarios, marine organisms experience reduced contrast and color-cues that may be used for survival (for a wonderful review, see (Caves and Johnsen, 2021). In cases of high turbidity and bathymetric shifts optical properties are similar; both entail reduced illumination and narrowing of spectral breadth. Turbidity, however, can be dynamic and transient while bathymetric conditions are more stable; how might sensory adaptation across these timescales differ? What may be the impact of inhabiting turbid or bathymetric waterscapes on visually-mediated behavioral circuits?

## *Chemosensory Systems*

Physiological impacts of environmental pollutants are moderately well studied in terms of toxicology and human health, but less so in the context of the sensory ecology and neuroethology. In humans, multifactorial environmental pollution has been associated with poor health outcomes, including neurological disorders (Manisalidis et al., 2020), in addition to impacts on stress (Thomson et al., 2019) and immunity (Bauer et al., 2012; Thomson et al., 2019). Unsurprisingly, chemical pollutants can impair chemoreception in animals (Ajmani et al., 2016; Troyer and Turner, 2015). This is best described in aquatic species, as opposed to terrestrial species, where the olfactory epithelium is in direct contact with environmental pollutants in water (Whitlock and Palominos, 2022).

Ocean acidification impairs behaviors that rely on chemosensation, such as habitat settlement, predator avoidance, foraging, and others (for a review, see Leduc et al., 2013), though these studies rarely conduct individual sensory manipulations in isolation. High $CO_2$, and consequently, low pH, are proposed to modify the structure of odorants as well as odorant-receptor affinity, resulting in reduced olfactory sensitivity (Porteus et al., 2021). Brain homogenates of larval sea bream reared in acidified conditions displayed a pH-dependent reduction in GABA and acetylcholine neurotransmitter concentrations as well as changes in expression of genes related to olfactory transduction. Specifically, experimenters observed a downregulation of genes encoding positive gene regulators (the olfaction-specific G protein, $G_{olf}$, and G-protein signaling 2, *RGS2*), and opposingly, an upregulation in genes encoding for negative regulators (G protein-coupled receptor kinase, *GRK*, and *arrestin)*, suggesting that ocean acidification has direct influence on chemosensory transduction pathways, down to the molecular level (Jiahuan et al., 2018).

In addition to pollutants and ocean acidification, nano and micro-plastics also affect animal chemosensation. Breakdown of phytoplankton and marine algae during zooplankton grazing releases dimethyl sulfide. Dimethyl sulfide attracts seabirds to these zooplankton grazing sites in the midst of vast stretches of featureless ocean, effectively serving as an olfactory cue to inform seabirds of zooplankton prey location. Nano-and micro-plastics accumulate phytoplankton and marine algae on their surfaces, which also release dimethyl sulfide when decomposing; no zooplankton needed. Since dimethyl sulfide is released by decomposing phytoplankton and algae on microplastic surfaces, seabirds mistakenly consume microplastics (Savoca and Nevitt, 2014; Savoca et al., 2016). The same sensory cue seabirds evolved to reliably detect, localize, and depend on for their survival now serves as an anthropogenic sensory trap.

Physiological changes in response to anthropogenic stresses are slowly beginning to be described but underlying genetic or cellular systems driving their change are not well understood, as demonstrated by the ocean acidification example above. In a beautiful set of studies, Thimmegowda and colleagues (Thimmegowda et al., 2020) tracked activity of Giant Asian honeybees in distinct locations with varying levels of air pollution across Bangalore, India over the course of three years. High-levels of air pollution correlated with changes in honeybee survival, flower visitation, heart rate, and bee morphology. The experimenters found changes in gene expression related to metabolism, stress, and immunity in antennae and heart tissue. Intriguingly, lab-reared *Drosophila* exposed to the same field sites also underwent similar changes in physiology and gene expression, providing accessibility into the mechanistic consequences underlying human-generated environmental stress. Perhaps, there is an underlying sensory component to olfactory signaling that is impaired by certain pollutants, alongside other physiological processes.

Pollutants can also affect sensory processes outside of chemosensation through neurotoxic effects. A recent study revealed that chronic sublethal exposure to cholinergic pesticides impairs wide-field visual motion detection in honeybees, as assayed through the optomotor response (Parkinson et al., 2022). This impairs the ability for honeybees to fly in straight, directed trajectories when they have gone off-course. At a molecular level, pesticide exposure resulted in altered stress and detoxification gene expression in the brain. Exposure to sulfoxaflor, one of the cholinergic insecticides tested, resulted in increased apoptosis primarily in the optic lobe where visual information is processed. These results suggest that cholinergic pesticides impair encoding of optic flow cues and result in disrupted optomotor behaviors.

Temperature, described in depth in the next section, also produces synergistic effects with pollutants that exacerbates their neurotoxic impacts, potentially by increasing contaminant uptake and/or modifying bioavailability (Sokolova and Lannig, 2008). Polystyrene plastic nanoparticle exposure, for example, exerts changes in specific metabolic pathways that lead to brain tissue degradation and disruption in circadian rhythms in zebrafish. In this study, spectral analyses also detected polystyrene particles in brain homogenates at all experimental temperatures tested but were generally found to be more concentrated at higher temperatures; any level of microplastic exposure is likely detrimental, but even more so at higher temperatures (Sulukan et al., 2022). Sodium cyanide poisoning from illegal, yet still prevalent, cyanide fishing practices also poses a growing threat to reef fish nervous systems due to the synergistic impacts from warming oceans (Madeira et al., 2020). Additional work is needed to understand the effects of combined perturbations on nervous system function.

*Other sensory systems*

Sensory systems outside of auditory, visual, or chemosensory modalities are also impacted by anthropogenic activity. Magnetic fields created by electric current flow in subsea cables generate electric fields between 0.5 and 100 µV/m (England and Robert, 2022), which is within the detection range of aquatic electroreceptive organisms (Hutchison et al., 2020; Peters et al., 2007). Experimentally, magneto-receptive lobsters and electro-sensitive skates exhibit increased exploratory behavior when exposed to electromagnetic fields from simulated underwater cable deployments (Hutchison et al., 2020). Deeper descriptions of electromagnetic sensing are needed in order to understand how these increases in exploratory behavior may functionally affect electro- and magnetoreceptive sensory systems.

It is intriguing to consider how sensory processing may be modified in cases of environmental perturbation, where the cause is indirect or direct; could compensation by alternate sensory systems result in the same efficacy and robustness in behavior? How might the brain know when to produce behavioral or genetic adaptations in response to inadequate sensory cues? Within a neural-circuit context, how might circuit elements compensate for rapid change? How might the *how* and *why* of sensory ecology change? Anthropogenic activity is essentially creating a *new* stimulus set; mechanistic understanding of sensory system biology is tractable in a lab setting if utilizing environmentally informed stimuli.

# Temperature

A central concern of anthropogenic environmental impacts is temperature fluctuation that arises from elevated greenhouse gases which trap heat in the atmosphere. Climate model simulations predict that the global average temperature will increase between 1.1 °C and 5.4 °C by the year 2100 (Friedlingstein and Prentice, 2010; Markus Meier, 2012), depending on the extent and efficacy of climate action. Temperature itself has profound impacts on a multitude of physiological processes, for example sex determination (Rhen and Schroeder, 2017) and metabolism (Schulte, 2015). Likewise, temperature alters biophysical properties of cells such as membrane fluidity (Fan and Evans, 2015) and ion permeability (Lamas et al., 2019).

In neurons, even slight physiological fluctuations in temperature alter neural physiology and function (Moser et al., 1993; Andersen and Moser, 1995). Minor reductions in temperature have dramatic impacts on synaptic vesicle release processes and presynaptic mechanisms of neural plasticity (Katz and Miledi, 1965; Micheva and Smith, 2005). Elevated temperatures enhance neural excitability by altering intrinsic membrane properties (Kim and Connors, 2012; Buzatu, 2009). More specifically, ion channels and signaling proteins are temperature sensitive (Tang et al., 2010; Robertson and Money, 2012) such that elevated temperatures result in changes to the resting membrane potential, the generation of action potentials, the velocity of action potential propagation, and action potential duration (Buzatu, 2009). These alterations in membrane excitability are reflected in unstructured neural circuit activity (see examples below) and have been shown to modify a range of neurobiological processes including sensory transduction, motor reflexes, locomotion, learning, and memory (Vornanen, 2017). Mechanisms of thermal acclimation counteract effects of acute elevated temperatures by changing the number of ion channels and through differential expression of temperature-specific isoforms of ion channels (Vornanen, 2017). Biology displays no shortage of temperature-dependent processes. An emerging field of research seeks to understand how temperature change arising from climate change influences neurobiological function and the resilience of neural circuits.

This is perhaps best studied in marine invertebrates where water temperature varies dramatically daily and seasonally. Body temperature of ectothermic organisms track ambient temperatures so they must rely on rapid behavioral and physiological adaptations for survival. In normal conditions ion flow is balanced across neuronal membranes, allowing neural circuits to produce appropriate responses and activity patterns. In the stomatogastric nervous system (STNS) of the crab, for example, ion channel flow is modulated to maintain function of the rhythmic circuit that mediates pyloric filtering of food in the stomach, despite variability in temperature (Tang et al., 2010b; Soofi et al., 2014; Kushinsky et al., 2019). Pattern generating neurons in the crab are remarkably robust to changes in temperature up to a point, however above this limit, which is regulated by adaptive homeostatic mechanisms (Tang et al., 2012), temperature causes an imbalance of inhibitory and excitatory conductances leading to membrane leakage and 'crashing,' or dysfunction of the rhythmic circuit by shunting action potential generation (Stein and Harzsch, 2021). Interestingly, neuromodulatory input enhances the robustness of STNS rhythms (Städele et al., 2015; Haddad and Marder, 2018; DeMaegd and Stein, 2021) which may also contribute to robustness of temperature adaptation.

Using wild-caught crabs, an inadvertently longitudinal study discovered that crash temperatures at which the crab STNS circuit ceased to function varied across years such that higher crash temperatures of STNS neurons reflected increases in ocean temperature (Marder and Rue, 2021). This suggests that the robustness of this circuit relies on the ability to acclimate to ambient temperatures, but begs the question of how robust this adaptive capacity will fare under continually warming oceans; in other words, what are the limits? What features are associated with enhanced resilience and adaptability in individuals and how might this contribute to behavioral adaptation, migration, or attrition?

Learning and cognitive function also suffer temperature-dependent impairment in humans (Taylor et al., 2016; Khan et al., 2021; Park et al., 2021) and in animals (reviewed in Soravia et al., 2021). Zebrafish exposed to temperatures 2°C above their voluntary thermal maximum (the temperature that elicits avoidant behavior; Rey et al., 2015 for a prolonged period, exhibited reduced interest in novel environments and impaired cognitive performance in Y-maze behavioral tests that correlated with a downregulation of proteins associated with synaptic transmission(Toni et al., 2019). Similar impairments and related patterns in gene expression from acute extreme temperatures have been replicated across many species (Soravia et al., 2021). It is proposed that ongoing adaptation of heat shock proteins will determine thermal tolerance of organisms (Hupało et al., 2018; González et al., 2016). In fact, at short timescales, heat shock protein expression instead prevents cellular damage and improves long memory formation in great pond snails (Sunada et al., 2016)

In a fascinating study, Port Jackson sharks raised in predicted end of the century temperatures (3°C increase) out-performed their normal-temperature reared counterparts in a quantity discrimination task, in both time to learning and accuracy at the task (Pouca et al., 2019). While changes in metabolic demands may perhaps be the driver of these observations, there may be other neural correlates or processes associated with cognition. Cognition itself may act as a buffer to counteract detrimental impacts of changing ecosystems (Brown, 2012). More work is needed to understand how temperature-dependent impairments in cognition are preserved or unprotected by compensatory and adaptive mechanisms, and how cognition may play a role in determining species-specific responses to changing ecosystems.

# Discussion

Rapid anthropogenic environmental change is occurring and consequently, its influence on animal behavior and corresponding neural substrates is a very large experiment-in-progress. While nervous systems evolved to adapt to environmental changes at non-disruptive scales, current change rates may outpace adaptive behavioral responses (Radchuk et al., 2019). The underlying impacts that changing environments have on neural processes are considerably underexplored, providing ample opportunity for scientific research with the potential for transformative discoveries.

Neuroscientists generally draw conclusions about the brain using experimentally generated, well-controlled physiological or environmental perturbations. In the natural world environmental perturbations, however, are variable, as are the physiologies, diets, genetic backgrounds, and other features of natural populations. Neuroscience as a field has gone great lengths to clamp these variables in exchange for reliability, and the subsequent ability to draw conclusions. Experimentally approaching neurobiology and changing ecosystems in such a way that captures true environmental variability while maintaining experimental control will not always be feasible, highlighting the necessity for a coalescence between lab and in-field studies. As a lab-only approach, experimenters can implement predictions from global climate models as controlled experimental variables, as demonstrated by multiple studies presented in this perspective piece. Exposure to in-field studies can begin by increased crosstalk and collaboration between ecologists, evolutionary biologists, and neurobiologists. Together, this will not only allow a deeper understanding of limits and capabilities of cell and neural circuit function, but help improve predictive models for climate change impact on species and populations.

Though many natural populations are threatened by anthropogenic activity, some species are thriving. Jellyfish, which have an increasing global ecological and economic significance, are more successful in warmer water, potentially due to longer breeding periods (Brotz et al., 2012). They have also recently been demonstrated as a genetically tractable model organism (Weissbourd et al., 2021). Comparative studies may allow us to understand what parameters enable some organisms to thrive while others fail to adapt.

Changes in seasonality, resource availability, and resource quality are shifting the locations that animals inhabit, leading to new inter-species interactions (Kharouba et al., 2018; Cohen et al., 2018). As new predator-prey interactions emerge, or new forms of inter-species cooperation develop, these imminent phenomenological observations may serve as a basis for fascinating novel studies of animal behavior and adaptive neural circuits. Conceivably, environmental pressures are creating new mechanisms that counteract stress that we have yet to discover.

There is growing advocacy for neuroscientists to take specific actions to mitigate the climate crisis by reducing emissions (e.g., virtual conference attendance, reduced travel, and reductions in waste and energy consumption in lab operations. For a guide, see (Aron et al., 2020). Similarly, this emerging area of scientific research may have societal implications that may help to shape climate adaptation, rather than mitigation, responses. In the case of the study of pollutants and pollinators (Thimmegowda et al., 2020), there are direct implications for food security, based on the chemosensory ecology of bees. Potential policy and conservation applications likewise can be derived or informed by neuroscientific studies related to anthropogenic impacts on animals and ecosystems. This new avenue may evolve into applied environmental neuroscience research driven by the underlying question of how nervous systems cope with increasingly variable and extreme changes in ecosystems. Anthropogenic impacts on ecosystems and natural animal populations are inescapable, but they provide a new challenge to discover fundamental properties of cells and neural circuits.

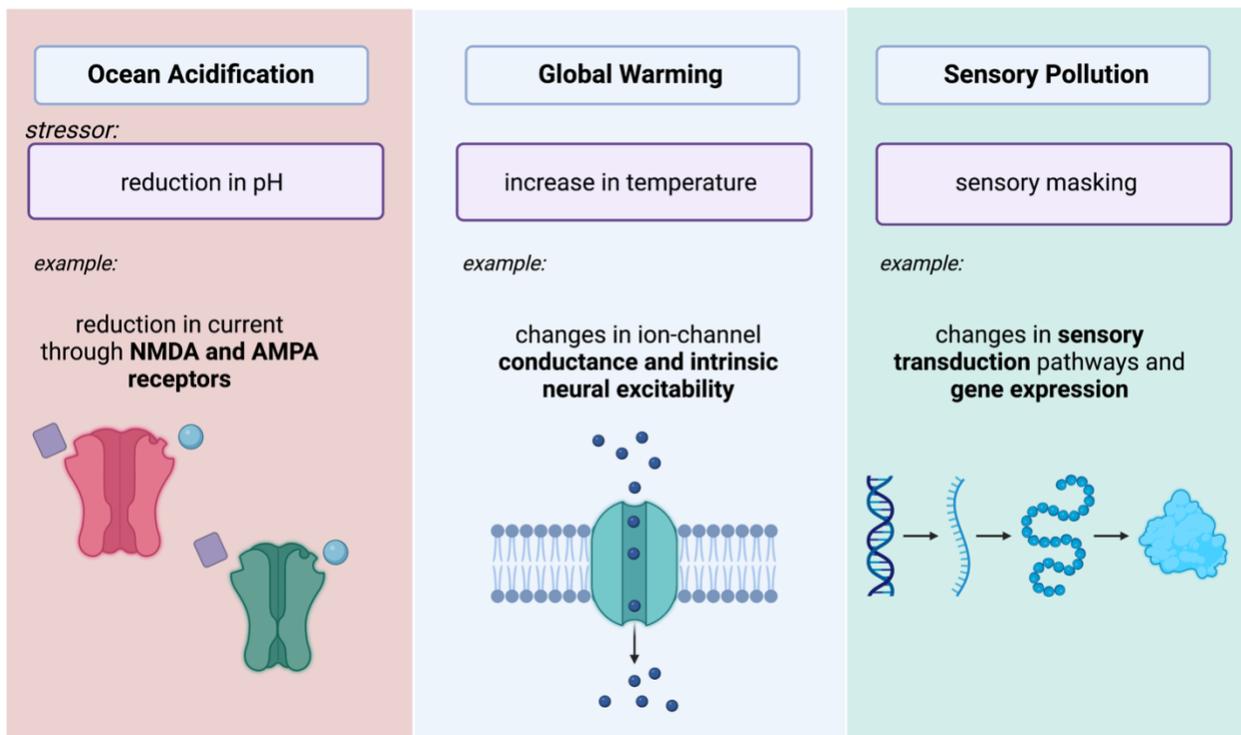

Figure 1: Summary schematic depicting three major changes in ecosystems that impose stressors on neural systems. Ocean acidification causes decreases in aquatic pH, potentially leading to reductions in current through NMDA and AMPA Receptors. Increasing temperatures impact intrinsic excitability of neurons. Changes in sensory environments due to sensory pollution (light, noise, and chemical pollution) may lead to changes in sensory transduction and gene expression. Image created with BioRender.

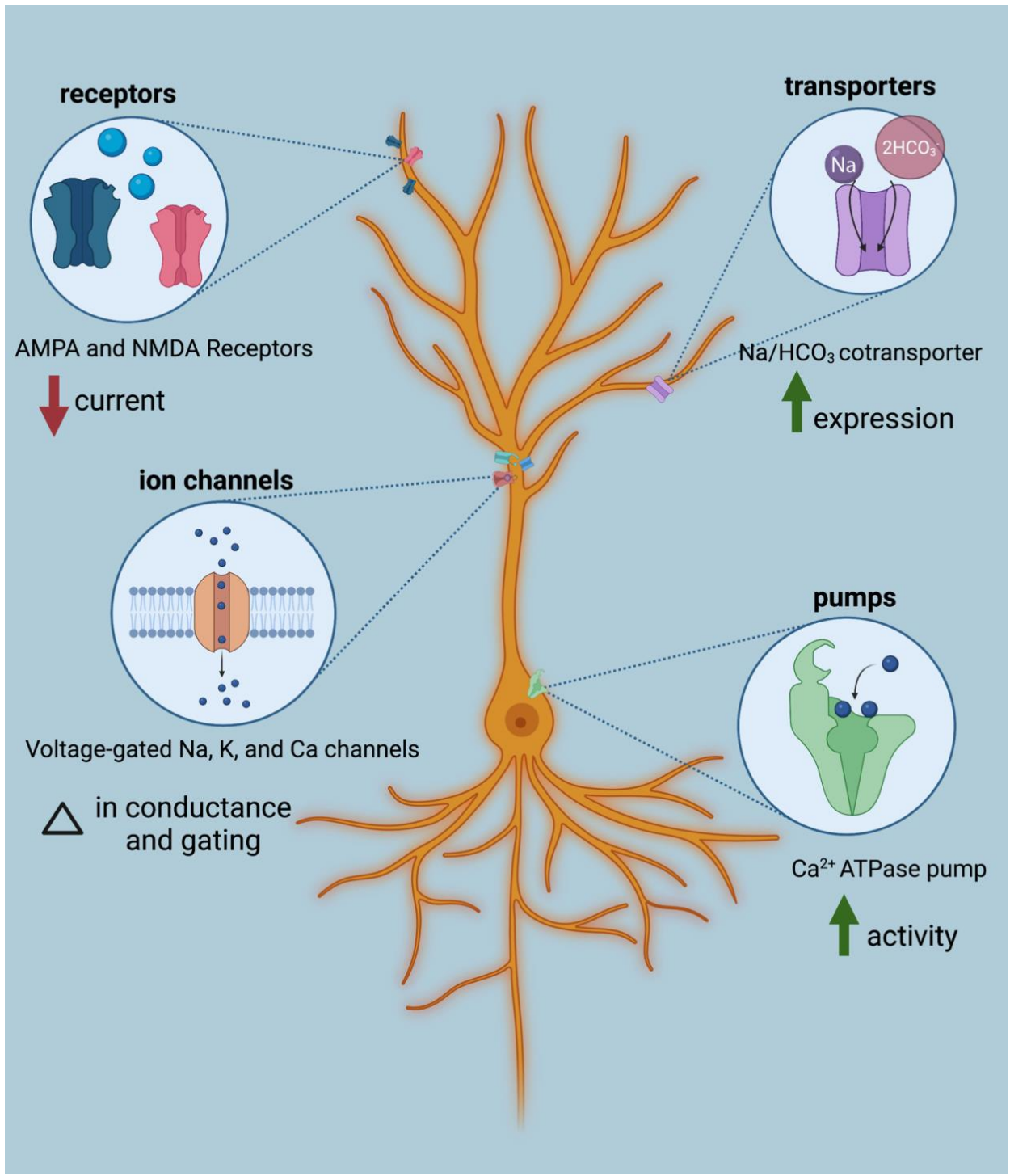

Figure 2: Classes of membrane proteins expressed in neurons are influenced by acidic conditions modulate neural excitability. Shown is one example of each; for additional impacts of low pH on membrane protein function, see Ruffin et al., 2014. Image created with BioRender.

# Acknowledgements

We would like to acknowledge our colleagues at The Kavli Foundation for their input and guidance in shaping the ideas presented here. We would also like to acknowledge Dr. Philip Parker and Dr. Jennifer Hoy for their feedback on the manuscript, and Dr. Eve Marder whose work inspired the genesis of ideas that led to this review.